\documentclass[twocolumn,superscriptaddress,showpacs,preprintnumbers,amsmath,amssymb]{revtex4}

\usepackage{graphicx}
\usepackage{dcolumn}
\usepackage{bm}

\newcommand{\nc}{\newcommand}		
\nc{\vc}[1]	{\mbox{\boldmath $#1$}}	
\nc{\del}       {\partial}              
\nc{\bra}       {\langle}               
\nc{\ket}       {\rangle}               
\nc{\bras}[1]   {\langle #1|}           
\nc{\kets}[1]   {|#1\rangle}            
\nc{\mapleft}[1]{			
 \smash{\mathop{\,			%
  \hbox to 1.5cm{\rightarrowfill}\, }\limits_{#1}}}
\nc{\beq}     {\begin{eqnarray}}
\nc{\eeq}    {\end{eqnarray}}
\nc{\fra}     {\frac{1}{2}}
\nc{\nn}    {\\ \nonumber}
\nc{\bx}     {\bold x}
\nc{\be}    {\bold e}
\nc{\bE}    {\bold E}
\nc{\bB}    {\bold B}
\nc{\bl}    {\bold l}
\nc{\bS}    {\bold S}
\nc{\bs}    {\bold s}
\nc{\bA}    {\bold A}
\nc{\by} {\bold y}
\nc{\br} {\bold r}
\nc{\bv} {\bold v}
\nc{\bX} {\bold X}
\nc{\bR} {\bold R}
\nc{\bk} {\bold k}
\nc{\bz} {\bold z}
\nc{\vap} {\varphi}
\nc{\ti} {\tilde}
\nc{\lm} {\lambda}
\nc{\al} {\alpha}
\nc{\bt} {\beta}
\nc{\gm} {\gamma}
\nc{\dl} {\delta}
\nc{\tht} {\theta}
\nc{\sig} {\sigma}

\nc{\vs}      {\vspace{-0.275cm}}

\begin{document}
\preprint{APS/123-QED}

\title{Tensor-optimized few-body model for s-shell nuclei}

\author{K. Horii}
\email{horii@rcnp.osaka-u.ac.jp}
\affiliation{Research Center for Nuclear Physics (RCNP), Osaka University, Osaka 567-0047, Japan}
\author{H. Toki}
\email{toki@rcnp.osaka-u.ac.jp} 
\affiliation{Research Center for Nuclear Physics (RCNP), Osaka University, Osaka 567-0047, Japan}

\author{T. Myo}
\email{myo@ge.oit.ac.jp} 
\affiliation{Osaka Institute of Technology, Osaka, Osaka 535-8585, Japan}

\author{K. Ikeda}
\email{k-ikeda@riken.go.jp} 
\affiliation{RIKEN Nishina Center, 2-1 Hirosawa, Wako, Saitama 351-0198, Japan}

\date{\today}
             
\vspace{0.5cm}

\begin{abstract}
As a new scheme of treating the tensor interaction of the nucleon-nucleon interaction, there is a proposal of a tensor-optimized shell-model (TOSM) for the study of medium and heavy nuclei.  The TOSM includes the deuteron-like tensor correlation and provides quite a good reproduction of the binding energy and the size of $^4$He as compared with rigorous few body calculations.  We propose a tensor-optimized few-body model (TOFM) using the spirit of the TOSM approximation in the few body framework with bare nucleon-nucleon interaction.  We find that the TOFM can account for the strength of the tensor interaction very well and almost reproduces the full energy and various energy components as compared with rigorous few body calculations. 
\end{abstract}


\maketitle

It is very important to describe nuclear structure by using bare nucleon-nucleon (NN) interaction.  Such an effort has been made for a few body~\cite{kamada01} and light mass nuclei up to about A$\sim$12~\cite{peiper01}.  The difficulty of solving nuclear ground states by using the NN interaction is the presence of the strong short range repulsion and the medium range tensor interaction caused by the pion exchange.  The calculated results on the deuteron indicate that the s-wave component has a distinct dip at the short distance and the probability of the d-wave component is of order of 5\%.  Although the d-wave component is small, the dominant attraction of order of 80\% is caused by the tensor interaction to provide the small binding energy by canceling with a large kinetic energy for the case of the AV8' interaction.  These features have to be handled in any theoretical frameworks for a quantitative account of nuclear states.  The calculated results of few body frameworks provide extremely a good account of nuclear ground states and a few excited states by including a phenomenological three body interaction~\cite{peiper01}.


In recent years, two very important methods were proposed for the description of finite nuclei.  One is the unitary correlation operator method (UCOM) and the other is the tensor optimized shell model (TOSM).  The UCOM is a method to handle the short range correlation in terms of the unitary correlation operator by truncating the resulting many body operators at the 2 body level~\cite{feldmeier98}.  The TOSM is a method to treat the strong tensor interaction in the 2p-2h model space in the shell model basis~\cite{myo09}.   Hence, we are able to calculate medium and heavy nuclei by using bare NN interaction.  We discuss here the essence of the TOSM approximation.  The strong tensor interaction acting on two nucleons in a spin-saturated shell model state oughts to excite them into two particle states.  Hence, it is a minimum requirement to take 2p-2h excitations in order to treat the tensor interaction, which provides a large attraction to form nucleus.  Another important feature is to use various range (short range in particular) gaussian wave functions so that the variational function is able to describe the strong tensor correlation.  Myo et al. introduced the TOSM in order to describe the spin-orbit splitting effect in $^5$He due to the blocking mechanism of the $p_{1/2}$ orbit caused by the deuteron-like tensor correlation~\cite{myo05}.  Another important application of the TOSM was the formation of the halo structure in $^{11}$Li~\cite{myo072}.  Again the tensor correlation is responsible to disfavor a configuration of 2 neutrons in the $p_{1/2}$ orbit and makes the contribution of 2 neutrons in the $s_{1/2}$ orbit large to be the source of the halo structure.

A theoretical study was then performed by Myo et al. to use a bare NN interaction in the TOSM for $^4$He and the numerical results were compared with those of few body methods~\cite{myo09}.  They used the central correlation part of the UCOM in order to treat the short range repulsion in the NN interaction.  The comparison was quantitatively very successful.  The tensor matrix element comes out to be about 80\% of that of few body methods.  The binding energy was about 3 MeV smaller than that of the few body methods.   Hence, the TOSM can be used for the discussion of nuclear structure of medium and heavy nuclei, in particular for the study of the effect of the tensor interaction for various observables in those nuclei.   It is then very important to know from where the difference between the TOSM and the rigorous calculations comes from and what should be done in order to improve the TOSM description of nuclei.

In a few body systems, we have a powerful few-body technique to describe light nuclei by using the relative coordinates, which correspond to the coordinate of the nucleon-nucleon interaction.  It is much easier to take into account the tensor interaction and the short range repulsion.  On the other hand, the necessary anti-symmetrization of all the particle coordinates prohibits the application of the few-body method to larger mass nuclei.  One of the reasons of the difficulty is a large variational model space in which the whole energy is minimized.  Hence, it may be a good idea to introduce the spirit of the TOSM approximation in the few body framework.  This idea corresponds to take the s-wave configuration as the basis and add a single d-wave in a relative coordinate and do not take double and/or triple d-wave configurations.  The purpose of the present study is then to develop a tensor-optimized few-body model using the spirit of the TOSM and compare with results of rigorous calculations in a few body framework and to obtain a good approximate method to study larger mass nuclei.  

We state first the tensor optimized shell model (TOSM), which was developed for the description of the tensor interaction in the shell model basis.  The TOSM wave function is written as
\beq
|\Psi\ket=C_0 |0\ket+\sum_\alpha C_\alpha |2p-2h:\alpha\ket~.
\eeq
Here, $|0\ket$ denotes the 0p-0h shell model state.  The 2p-2h states $|2p-2h:\alpha\ket$ are excited by the operation of the tensor operator $S_{12}(\hat x) \propto [Y_2(\hat x)\times [\sigma_1\times \sigma_2]_2]_0$ on two particles in the shell model 0p-0h state.  In the tensor operator, $\hat x$ in the spherical harmonics function $Y_2$ denotes an angle of relative coordinate, and $\sigma$'s are the spin operators of two particles.  Hence, the two components are connected by the tensor operator as
\beq
\bra 2p-2h:\alpha|S_{12}(\hat x)|0\ket \ne 0~,
\eeq
where the quantum number $\alpha$ denotes all the possible 2p-2h states connected by the tensor operator.  The 2p-2h states are constructed to treat the tensor correlation by taking various gaussian range wave functions, particularly those with small size to optimize the role of the tensor interaction.  Hence, the essential approximation of the TOSM is the inclusion of the variational states, which are excited by only one operation of the tensor interaction.  As for the short range correlation, they use the central correlation part of the UCOM~\cite{feldmeier98} in the TOSM framework, because it is difficult to express the short range behavior in the shell model basis~\cite{myo09}.

On the other hand, a few body method uses the relative coordinates and hence it is more efficient to express the tensor and short range correlations, since the tensor interaction and the short range repulsion are expressed by the relative coordinates.  We only have to introduce all the states connected by one operation of the tensor interaction and take enough gaussian functions to express both the tensor and short range correlations.  We should end up introducing the following tensor-optimized wave function for $A=2, 3$ and 4 body systems by taking the essential features of rigorous calculations in the few body method.  Hence, we write a few body wave function as a linear combination of $S$ and $D$ wave components,
\beq
|\Psi \ket=|\Psi_{S}\ket+|\Psi_{D}\ket~.
\eeq
The condition of the D-wave component should be
\beq
\bra \Psi_{D}|S_{12}(\hat x)|\Psi_{S}\ket \ne 0~.
\eeq
Hence, $|\Psi_D\ket$ should contain $Y_2$-function only once.  Particularly the most essential D-wave state is to introduce a $Y_2$-function in the Jacobi coordinate $\bold x_1=\bold r_1-\bold r_2$ and perform all the necessary permutations so that all the particle pairs profit the use of the strong tensor interaction.  

The S-wave component for s-shell nuclei with $A\leq 4$ is written as
\beq
\label{swavecomp}
|\Psi_{S}\ket=\sum^{N_S} C_S^B\mathcal A|\psi_0(\bold B_S,\{x_i\}) \chi_s(\{\sigma_i\}) \chi_t(\{\tau_i\})\ket~.
\eeq
The spin wave function with total spin $s$ for a A=4 system is written as
\beq
\label{spin}
\chi_s(\{\sigma_i\})&=&[[[\chi_\fra(\sigma_1)\times\chi_\fra(\sigma_2)]_{s_{12}}\times \chi_\fra(\sigma_3)]_{s_{123}}\nn&&\times \chi_\fra(\sigma_4)]_s~.
\eeq
The iso-spin wave function is written similarly as
\beq
\label{isospin}
\chi_t(\{\tau_i\})&=&[[[\chi_\fra(\tau_1)\times\chi_\fra(\tau_2)]_{t_{12}}\times \chi_\fra(\tau_3)]_{t_{123}}\nn&&\times \chi_\fra(\tau_4)]_t~.
\eeq
On the other hand, the spatial wave function is written as \cite{varga97}
\beq
\label{swave}
\psi_0(\bold B_S,\{x_i\})=\exp(-\fra \ti \bx \bold B_S \bx )
\eeq
Here, $\bx$ represents the relative coordinate vector $\bx=(\bx_1,\bx_2,..)$  and $\ti \bx \bold B_S \bx$ means the short-hand notation of $\sum_{ij}^{N_P-1} B_S^{ij} \bx_i \cdot \bx_j$.  The relative coordinate vectors $\bx_i$ with $i=1,..,N_P-1$ are expressed in terms of the particle coordinates $\br_j$ with $j=1,..,N_P$, where $N_P$ is the number of particles of a nucleus.  
This variational function is able to express the short range correlation.  

As for the D-wave component, we have one $Y_{2M}$ function in the spatial wave function.
\beq
\label{dwave}
\psi_{2M}(\bold B_D,\{x_i\})= \exp(-\fra \ti \bx \bold B_D \bx )|\ti u\bx|^2 Y_{2M}(\ti u \bx)
\eeq
Here, the global vector $\ti u \bx$ is defined as the linear combination of the vector $\bx$ as $\ti u \bx=\sum_i^{N_P-1} u_i \bx_i$.  The gaussian wave function with the function $\ti \bx \bold B_D \bx=\sum_{ij}^{N_P-1}B_D^{ij} \bx_i \cdot \bx_j$ is able to optimize the role of the tensor interaction with small range gaussian functions in the d-wave part as the case of the TOSM.  Since the spatial wave function now has a finite angular momentum, we have to make coupling of the spatial wave function with the spin wave function to get a desired total spin $s$.
\beq
\label{dwavecomp}
|\Psi_{D}\ket=\sum^{N_D} C_D^B\mathcal A| [\psi_2(\bold B_D,\{x_i\})  \chi_{s'}(\{\sigma_i\})]_s \chi_t(\{\tau_i\})\ket
\eeq

We should know the role of the antisymmetrization operator $\mathcal A$, which is defined as
$
\mathcal A =\sum_{i=1}^{N!} p_i \mathcal P_i
$.
It consists of the particle exchange operator with the phase $p_i$ due to even or odd permutation, where the operator $\mathcal P_i$ makes the transformation of the relative coordinate $\bx_P=\mathcal T_i \bx$, the exponent of the correlated gaussian basis function and the generalized coordinate in the spherical harmonics takes the form after the antisymmetrization of the particles 
as $\bold B_P=\ti {\mathcal T} \bold B \mathcal T$ and $u_P= \ti {\mathcal T} u$.  The spin and isospin wave functions are subject to the antisymmetrization also.  Thus, the anti-symmetrized wave function contains quite a general expression of wave functions and the manipulation of various matrix elements is highly involved.  We introduce the representation of the global vector for the basis wave functions in the same way as the Niigata group~\cite{suzuki08}.  In the stochastic variational method (SVM), the gaussian ranges $\bold B_S$ and $\bold B_D$ are generated randomly and choose the most suitable wave functions successively~\cite{varga97}.

We obtain the wave function and energy in a few-body framework by taking minimization of the total energy $\delta \frac{\bra \Psi|H| \Psi \ket}{\bra \Psi| \Psi \ket}=0$.  The ranges of the spatial wave functions $\bold B_S$ in Eq. (\ref{swave}) and $\bold B_D$ in Eq. (\ref{dwave}) and their amplitudes $C_S^B$ and $C_D^B$ are chosen variationally.  We may take also the intermediate spin and isospin quantum numbers in Eqs. (\ref{spin}), (\ref{isospin}) and (\ref{dwavecomp}) with constraint that the final spin and isospin are $s$ and $t$.  Although there are several few-body methods, we work out numerical calculations using the SVM of the Niigata group~\cite{suzuki08}.   We refer all the details of calculations in Ref. \cite{suzuki08}.  

\begin{figure}
\centering
\includegraphics[width=8.0cm,clip]{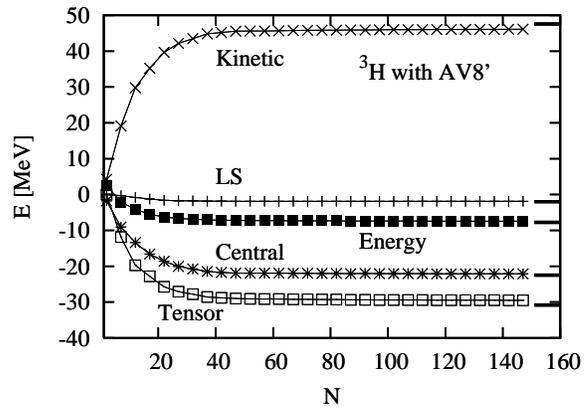}
\caption{Various energy components in the triton in the TOFM with AV8' as functions of the number $N=N_S+N_D$ of the basis wave functions.  Those energy components are the kinetic energy denoted by cross, the spin-orbit energy denoted by plus, the central interaction energy denoted by plus-cross, the tensor interaction energy denoted by open square and the total energy denoted by closed square.  The corresponding energy components obtained by rigorous calculations are shown in the right hand side.}
\label{fig:3h}
\end{figure}

\begin{figure}
\centering
\includegraphics[width=8.0cm,clip]{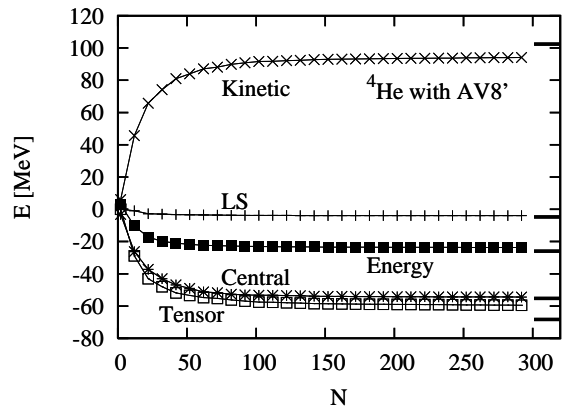}
\caption{Various energy components in $^4$He in the TOFM with AV8' as functions of the number of basis wave functions.  The notation is the same as Fig.\ref{fig:3h}.}
\label{fig:4he}
\end{figure}

We show numerical results in the TOFM and compare with the rigorous few body calculation for A=2, 3 and 4 systems and the TOSM calculation of Myo et al. for $^4$He~\cite{myo09}.  The results for the deuteron shown in Table I and II indicate the importance of the tensor interaction.  We work out numerical calculations with the TOFM using the AV8' potential for A=3 and 4.  We include only one relative coordinate with $l=2$ angular momentum state $\ti u \bx=\bx_1$, while the antisymmetrization $\mathcal A$ takes care of all the necessary permutations.  We base our numerical calculation on the computer code of Varga and Suzuki~\cite{varga97} and add the tensor component in the variational wave function.  We obtain variationally suitable gaussian ranges $\bold B_S$ and $\bold B_D$ and their amplitudes $C_S^B$ in Eq. (\ref{swavecomp}) and $C_D^B$ in Eq. (\ref{dwavecomp}) by diagonalization.  In Fig.\ref{fig:3h}, we show the energy convergence for $^{3}$H as functions of the number $N$ of the basis wave functions. The energy components and the total energy are shown up to $N=N_S+N_D=$150.   We see a good convergence is achieved already around $N$=50.  All the energies are compared with the SVM calculation shown in the right hand side of the figure.
We show in Fig.\ref{fig:4he} the case of $^4$He, which is obtained by using the $l$=2 component with $\ti u \bx=\bx_1$ only.   In this case the convergence is slow and we ought to go up to about $N$=300. The results of the rigorous few-body calculations for $^4$He are shown in the right hand side of this figure.  Although the convergence is achieved for the central and the spin-orbit energies, the tensor interaction energy and also the kinetic energy are not yet converged.

We show in Table I the total energy and various energy components as the kinetic energy, the central interaction energy, the tensor interaction energy and the spin-orbit interaction energy for the deuteron, $^3$H and $^4$He.  We do not include the Coulomb energy in this comparison.  Although very small, we add other $l=2$ components ($\tilde u \bx=\bx_2$ and $\bx_3$) in the calculation for this Table.  The present result is compared with the rigorous SVM calculation by Suzuki et al.~\cite{suzuki08}. The results of various few body methods are essentially the same as the SVM result~\cite{kamada01}.  As for $^4$He, we also compare the results with the TOSM.

We would like to discuss the comparison of the TOFM with the SVM  for $^3$H and $^4$He.  The TOFM results are compared almost perfectly with the SVM results for $^3$H.   For $^4$He, the binding energy is -24.08 MeV, which is close to the total energy of -25.92MeV of the SVM.  The central interaction energy and the spin-orbit energy are quite close to those of the SVM.  On the other hand, the tensor interaction energy is about 8MeV smaller and the kinetic energy is about 7MeV smaller than the SVM result.  Hence, the tensor component and the kinetic energy are slightly underestimated.
\begin{table}
\caption{\label{para} Various energy components with AV8'.}
\begin{center}
\begin{tabular}{cccccc}
\hline\hline
Nucleus   &~Energy~& ~Kinetic~&~Central~&~Tensor~&~ LS~\\  
\hline
deuteron~~ & ~-2.23~&~19.95~&~ -4.49~&~-16.64~&~-1.03~\\
\hline
$^3$H(TOFM)~~&~-7.54~&~46.67~&~-21.98~&~-30.47~&~-1.95~\\
SVM\cite{suzuki08} &~-7.76~&~47.57~&~-22.49~&~-30.84~&~-2.00~\\
 \hline
$^4$He(TOFM)~~&~-24.08~&~95.53~&~-54.61~&~-60.95~&~-4.05~ \\
TOSM\cite{myo09}&~-22.30~&~90.50~&~-55.71~&~-54.55~&~-2.53~ \\
SVM\cite{kamada01} &~-25.92~&102.35~&~-55.23~&~-68.32~&-4.71~\\
\hline\hline
\end{tabular}
\end{center}
\end{table}

\begin{table}
\caption{\label{para} Various energy components with G3RS.}
\begin{center}
\begin{tabular}{cccccc}
\hline\hline
Nucleus   &~Energy~& ~Kinetic~&~Central~&~Tensor~&~ LS~\\  
\hline
deuteron~~ & ~-2.28~&~16.48~&~ -7.29~&~-11.46~&~-0.00~\\
\hline
$^3$H(TOFM)
&~-7.61~&~39.82~&~-26.70~&~-20.69~&~-0.04~\\
SVM\cite{suzuki08} 
~~&~-7.73~& ~40.24~&~-26.80~&~-21.13~&~-0.03~\\
 \hline
$^4$He(TOFM)
&~-25.22~&~84.83~&~-66.21~&~-43.66~&~-0.17~\\
SVM\cite{suzuki08} 
~~&~-26.05~&~86.93~&~-66.24~&~-46.62~&~ -0.13~ \\
\hline\hline
\end{tabular}
\end{center}
\end{table}
As for the comparison with the TOSM calculation for $^4$He, the present calculation of the energy value is better than the TOSM result~\cite{myo09}.  Large differences are found in the matrix elements of the kinetic energy and the tensor interaction.  This difference should mean that the TOSM calculation can be improved by taking a more general UCOM correlation function.   Some defect in the short range correlation may be seen also in the LS component, which is underestimated in the TOSM.  

In order to see the interaction dependence, we show the case of the G3RS bare NN-interaction in Table II~\cite{tamagaki68}.  In this case, the tensor interaction is weaker than the case of the AV8' potential.  Hence, now the agreement of the TOFM with the full calculation is impressive.  We have calculated also the second 0$^+$ state in $^4$He.  The excitation energy comes out to be $E_X=18.5$MeV, which is to be compared with the experimental data of 20.21MeV.  The amount of the tensor energy for the $0^+_2$ is close to the one of $^3$H, which suggests the 3N+N structure.    Excited states are described nicely in the TOFM, while there is a Faddeev model approach~\cite{navratil99}.

We have formulated a tensor-optimized few-body model (TOFM) in the spirit of the TOSM.  The TOFM wave function contains the S-wave and D-wave components.  The D-wave component contains only one $Y_2$ component.  Hence, the TOFM approximation makes the variational space much smaller that the one of the rigorous calculation.  We have calculated $A=2, 3$ and 4 body systems in the TOFM and compared with the rigorous calculations.  As for $A=3$, the inclusion of single $Y_2$ component in the $\bold x_1$ coordinate provides essentially the same results as the full model space calculation.  This result indicates that one $Y_2$ component is enough to take care of the tensor interaction in the three body system.   As for $A=4$, we again obtain good reproduction of the rigorous results, but the TOFM slightly underestimates the tensor interaction with single $Y_2$ component.  The difference between the TOFM and the rigorous calculation comes essentially from the additional D-wave component in the other p-n pair.  The inclusion of two $Y_2$ functions in the variational space brings the total energy very close to the rigorous result with AV8' for $^4$He~\cite{hiyama03}.  

The present results indicate that nuclei like to have deuteron configurations and it is satisfactory to take the deuteron-like tensor correlations in the wave function.  The present study is very encouraging to extend our study for nuclei with $A\geq 5$ in the TOFM and to describe medium and heavy nuclei using the TOSM approximation with better description of the short range correlation.

The authors are grateful to Prof. H. Horiuchi and Prof. E. Hiyama for fruitful discussions and encouragements.  This work is supported by the JSPS grants 21540267 and 21740194.


\begin{thebibliography}{11}

\bibitem{kamada01} H. Kamada et al., Phys. Rev. C64 (2001) 044001, and references therein.

\bibitem{peiper01} S.C. Pieper and R.B. Wiringa, Annu. Rev. Nucl. Part. Sci. 51 (2001) 53.


\bibitem{feldmeier98}
H. Feldmeier, T. Neff, R. Roth and J. Schnack, Nucl. Phys. A632 (1998) 61;
T. Neff, and H. Feldmeier, Nucl. Phys. A713 (2003) 311.

\bibitem{myo09}
T. Myo, H. Toki and K. Ikeda, Prog. Theor. Phys. 121 (2009) 511. 

\bibitem{myo05} T. Myo, K. Kato and K. Ikeda,
Prog. Theor. Phys. 113 (2005) 763.


\bibitem{myo072} T. Myo, K. Kato, H. Toki and K. Ikeda,
Phys. Rev. C76 (2007) 024305.


\bibitem{suzuki08}  Y. Suzuki, W. Horiuchi, M. Orabi and K. Arai, Few Body System 42 (2008) 33.

\bibitem{varga97} K. Varga and Y. Suzuki, Comp. Phys. Comm. 106 (1997) 157.

\bibitem{suzuki00} Y. Suzuki and K. Varga, Lecture Note in Physics, 54 (1998) Springer, Berlin.

\bibitem{tamagaki68} R. Tamagaki, Prog. Theor. Phys. 39 (1968) 91.

\bibitem{navratil99} P. Navratil and B. R. Barrett, Phys. Rev. C59 (1999) 1906.

\bibitem{hiyama03} E. Hiyama, Y. Kino and M. Kamimura, Prog. Part. Nucl. Phys. 51 (2003) 223.




\end{thebibliography}
\end{document}